\documentclass[12pt]{iopart}

\usepackage{graphics}
\usepackage{graphicx}
\usepackage{xspace}
\usepackage{iopams}

\newcommand{\plainvar}{\sigma^2}
\newcommand{\var}[1]{\plainvar\left(#1\right)}
\newcommand{\ave}[1]{\left\langle #1 \right\rangle}
\newcommand{\Uvar}[1]{\plainvar_u\left(#1\right)}
\newcommand{\Uave}[1]{\left\langle #1 \right\rangle_u}
\newcommand{\Cvar}[1]{\plainvar_c\left(#1\right)}
\newcommand{\Cave}[1]{\left\langle #1 \right\rangle_c}
\newcommand{\UCvar}[1]{\plainvar_{u,c}\left(#1\right)}
\newcommand{\UCave}[1]{\left\langle #1 \right\rangle_{u,c}}

\newcommand{\FC}{\mathcal{F}}

\newcommand{\GC}{\mathcal{G}}

\newcommand{\gvec}[1]{\mathbf{#1}}

\newcommand{\rvec}{\gvec{r}}

\newcommand{\ddint}[1]{\mathchoice{\!\!\mathrm{d}^d#1\,}{\!\mathrm{d}^d#1\,}{\!\mathrm{d}^d#1\,}{\!\mathrm{d}^d#1\,}}

\newcommand{\elabel}[1]{\label{#1}}

\newcommand{\flabel}[1]{\label{fig:#1}}

\newcommand{\ie}{{\it i.e.}\xspace}

\newcommand{\activity}{\rho}
\newcommand{\UCsub}{{u,c}}
\newcommand{\probabAct}{a}

\begin{document}

\title{Block scaling in the directed percolation universality class}
\author{Gunnar Pruessner\dag\ddag}
\address{\dag Mathematics Institute, University of Warwick,
                    Gibbet Hill Road, Coventry CV4 7AL, United Kingdom}
\address{\ddag Present address: Department of Mathematics,
                    Imperial College London, 180 Queen's Gate, London
		    SW7 2AZ, United Kingdom}

\date{\today}
\ead{g.pruessner@imperial.ac.uk}

\submitto{NJP}

\pacs{05.70.Jk, 
      05.40.-a, 
      89.75.-k  
     }

\begin{abstract}
The universal behaviour of the directed percolation universality class
is well understood, both the critical scaling as well as finite size
scaling. This article focuses on the block (finite size) scaling of the
order parameter and its fluctuations, considering (sub-)blocks of linear
size $l$ in systems of linear size $L$.
The scaling depends on the choice of the ensemble, as only
the conditional ensemble produces the block-scaling behaviour as
established in equilibrium critical phenomena. The dependence on the
ensemble can be understood by an additional symmetry present in the
unconditional ensemble. 
The unconventional scaling found in the unconditional ensemble
is a reminder of the possibility that scaling functions themselves
have a power-law dependence on their arguments.
\end{abstract}

\maketitle

\section{Introduction}
The directed percolation (DP) universality class comprises a
huge number of non-equilibrium critical phenomena.  Janssen and Grassberger
\cite{Janssen:1981,Grassberger:1982} famously conjectured more than 25
years ago that under very general circumstances, all models with a
unique absorbing state belong to the DP universality class.  While this
conjecture has been confirmed numerically many times, 
evidence for the presence of the DP universality class in natural
systems
is still very scarce
\cite{Hinrichsen:2000b} (but see \cite{TakeuchiETAL:2007}).

One problem, when probing field data for the presence of DP is that
field data is more readily obtained in a single measurement rather than
as a
time series. However, the statistical features to be identified require
an entire ensemble of realisations of the process in question. Instead of using a
time series, one can resort to sub-sampling, \ie splitting a large sample
of size $L^d$ into $(L/l)^d$ small ones of size $l^d$. 
For example, a population pattern obtained by measuring the spatial
distribution of species could be split into several distinct blocks and
their mutual correlations analysed.

The question how the order parameter of an equilibrium system at the
critical point changes with the block size it is averaged over, has been
studied in great detail by Binder \cite{Binder:1981}. In
the present work, a corresponding analysis is applied to models belonging to the
DP universality class, more specifically to the contact process and to
directed percolation itself from the point of view of absorbing state
(AS) phase transitions. It turns out that the block averaged order parameter 
needs to be defined very carefully in order to reproduce standard finite
size scaling. In the
following, I will present numerical evidence and theoretical arguments
that block finite size scaling (FSS) in DP can be very different from what
is expected from equilibrium critical phenomena, 
depending on the choice of the ensemble. This observation can be readily
applied to the analysis of field data, and will be illustrated using
surrogate data.

\section{Method}
The order parameter of an absorbing state phase transition is the
activity $\activity$, which, in lattice models, is the density of active
sites. The activity vanishes as soon as the system hits the absorbing
state, from where it cannot escape. In a finite system, the absorbing
state is reached with finite probability from anywhere in phase space,
so that every (finite) system eventually becomes inactive,
$\lim_{t\to\infty}\activity(t)=0$, where $t$ measures the time in the
model. However, the order parameter $\activity$ signals a phase
transition in a temperature-like tuning parameter $p$, dividing the
parameter space into a region where the decay of the activity
$\activity$ is exponentially fast in time, from a region where for
sufficiently large systems it is practically impossible to observe
$\activity=0$. 

At least two (seemingly) different methods
have been devised to overcome the problem that strictly
$\lim_{t\to\infty}\activity(t)=0$ and
to obtain the phase transition even in finite systems:
One either introduces an external field which induces
activity in the system \cite{LuebeckHeger:2003} and analyses the model in the limit of
arbitrarily small fields, or one considers the
quasi-stationary state
\cite{OliveiraDickman:2005}. 
In this latter approach, after initialising and discarding a transient,
all averages are taken conditional to activity.
The activity entering the observables therefore never vanishes and
the order parameter is always non-zero, rendering for example moment ratios
well defined. One can show that both methods
produce asymptotically equivalent scaling results \cite{Pruessner:2008}.
In the present article, the quasi-stationary state was used to produce
individual samples.

Above the critical point $p_c$, \ie for $p>p_c$, the 
ensemble average of the activity $\activity$ in the thermodynamic limit
picks up as a power-law $\ave{\activity} = A ( p-p_c)^\beta$, where $A$ is
the amplitude, $p_c$ is the critical value of the tuning parameter and
$\beta$ is a universal critical index.  Similarly, fluctuations of
the order parameter scale as
$\var{\activity}\equiv\ave{\activity^2}-\ave{\activity}^2=B_{\pm} L^{-d} |
p-p_c|^{-\gamma}$, where $B_+$ and $B_-$ are the amplitudes above and
below the transition respectively, $L$ is the linear extent of the
system, $d$ is its spatial dimension and $\gamma$ is an independent
critical exponent. Both these power laws are asymptotes and thus acquire corrections
\cite{Wegner:72} away from the critical point.

Ordinary  FSS
\cite{FisherBarber:1972,Barber:83,PrivmanHohenbergAharony:1991} is
observed by tuning $p=p_c$ and considering the behaviour of the
observables as a function of the system size, $\ave{\activity}(L) = A'
L^{-\beta/\nu_\perp}$ and $\var{\activity}(L)=B' L^{-d+\gamma/\nu_\perp}$,
where $\nu_\perp$ is a third critical exponent. These exponents are
related by
$-d+\gamma/\nu_\perp=-2\beta/\nu_\perp$
\cite{MarroDickman:1999}, which in equilibrium corresponds to Josephson hyper-scaling
together with the Rushbrooke scaling law. 

Another kind of  FSS  can be explored in addition to the ordinary
one just described. Instead of considering the entire system,
observables are recorded within small blocks of linear extent $l$. To
this end, I introduce the observables $\Uave{\activity}(l;L)$ and
$\Uvar{\activity}(l;L)$, which are first and second cumulants of the activity
within those blocks of linear size $l$ in a system of size $L$. The
blocks are produced by dividing samples of linear size $L$ generated in the quasi-stationary
state into $(L/l)^d$ blocks of linear size $l$ each. This length $l$ is
conveniently chosen so that it divides $L$. By construction, each sample
of size $L^d$ contains at least one
active site and consequently at least one block of size $l^d$ has
non-vanishing activity. However, there may be up to $(L/l)^d-1$
inactive blocks.

In addition to these observables, I introduce $\Cave{\activity}(l;L)$ and
$\Cvar{\activity}(l;L)$, which are the first and second cumulant of the
activity conditional to activity within the respective block. That means
that inactive blocks are discarded when averaging. 
To
distinguish the two ensembles, the former (with cumulants
$\Uave{\activity}(l;L)$ and
$\Uvar{\activity}(l;L)$) will be called
``unconditional'' (subscript $u$) and the latter (with cumulants
$\Cave{\activity}(l;L)$ and
$\Cvar{\activity}(l;L)$) ``conditional''
(subscript $c$). 
One can derive the moments of the unconditional ensemble from the
corresponding moments in the conditional ensemble and vice versa by re-weighting,
because they differ only by a number of samples with vanishing
block-activity which are discarded in the conditional ensemble but not
in the unconditional one (see \cite{Pruessner:2008} for
similar considerations for the overall activity).
If the fraction of active blocks averaged over the entire unconditional
ensemble is $\probabAct(l;L)$, then the $n$th moment
of the unconditional ensemble $\Uave{\activity^n}$ is related to the $n$th
moment of the conditional ensemble $\Cave{\activity^n}$ by
\begin{equation}
\Uave{\activity^n}(l;L) = \probabAct(l;L) \Cave{\activity^n}(l;L)
\elabel{rel_uave_cave}
\end{equation}
for $n>0$.

More than 25 years ago, block scaling was investigated by Binder
\cite{Binder:1981} for the
Ising model, where the order parameter is
the magnetisation density $m$ rather than the activity $\activity$.
In these systems with a symmetric phase-space, there is no corresponding 
distinction of active and inactive blocks.
Na{\"i}vely transferring these
results to DP suggests 
\begin{equation}\elabel{oparam_scaling}
\UCave{\activity}(l;L)= C_\UCsub l^{-\beta/\nu_\perp} \GC_\UCsub(l/L)
\end{equation}
in leading order of $L$ and in the limit of $l$ being large compared to
a lower cutoff $l_0$, \ie $l\gg l_0$.  Below this constant threshold
$l_0$, \ie $l<l_0$, the order parameter deviates from the
behaviour predicted by \eref{oparam_scaling}. This phenomenon is known
and well understood in classical critical phenomena
\cite{StaufferAharony:1994,PrivmanHohenbergAharony:1991}. 
In the
following, data for $l\le l_0$ is not shown; in the two-dimensional contact
process and directed percolation $l_0$ was estimated to be $l_0\approx8$
and in
the one-dimensional contact process $l_0\approx 16$.  

The dimensionless
scaling functions $\GC_\UCsub(x)$ in \eref{oparam_scaling} are bounded from
above and usually also from below (away from 0) for all $x\in[0,1]$.
The scaling function is universal up to a pre-factor, which can always be
absorbed into the metric factor $C_\UCsub$. The latter is required for dimensional
consistency.

Scaling behaviour similar to \eref{oparam_scaling}
is expected to hold for higher order moments and
cumulants as well, in particular for the variance of the activity
\begin{equation}\elabel{var_box_scaling}
\UCvar{\activity}(l;L)= D_\UCsub l^{-2\beta/\nu_\perp} \FC_\UCsub(l/L) \ , 
\end{equation}
using the scaling relations cited above. Again, the scaling functions
$\FC_\UCsub$ are constraint by being bounded from above and (usually) away
from zero from below.

So far, I have described what is expected in absorbing state
systems as inferred from equilibrium critical phenomena. In the
remainder of this article, I will first introduce the non-equilibrium
models and the methods used
in the numerical simulations for this study. The results for the
different observables (activity, its variance and a moment ratio) in the
different ensembles (unconditional and conditional) are then discussed
in the light of analytical arguments. The article finishes with an
application to surrogate data, a
discussion of the implications and the wider context of the findings and
concludes with a brief summary.

\section{Results}
Most of the numerics in this work is based
on the contact process
\cite{Harris:1974,MarroDickman:1999} on a two-dimensional square
lattice with periodic boundary conditions, but the same results are found
for the one-dimensional contact process as well as for two-dimensional
site-directed percolation \cite{Grassberger:1989}. In fact, it will be argued that they are general
features of the directed percolation universality class, if not of all
absorbing state phase transition. 
In the two-dimensional contact process occupied sites turn empty
with extinction rate $1$, while empty sites become occupied with rate
$z \lambda$ with $z$ being the fraction of occupied nearest neighbours.
The critical value of $\lambda$ has been estimated numerically with
great accuracy 
$\lambda_c=1.64877(3)$
\cite{Dickman:1999} (I used $\lambda=1.6488$ in the present study).
The time scale is set by the extinction rate. The two-dimensional 
contact process belongs to the
DP universality class which is characterised by exponents $\beta=0.583(3)$,
$\gamma=0.297(2)$ 
and
$\nu_\perp=0.733(4)$ \cite{GrassbergerZhang:1996}, so that
$\beta/\nu_\perp=0.795(6)$.

In site-directed percolation in $2+1$ dimensions (BCC lattice) the time evolves
discretely \cite{Grassberger:1989,LuebeckWillmann:2005}. A site is
occupied in the following time step with probability $p$ if at least one
of its directed neighbours is occupied, otherwise it is empty. The
directed neighbours of a site are four sites in the preceding time step:
The site itself, its right and upper nearest neighbour and its upper
right next nearest neighbour.  The critical value of $p$ in this model
has been estimated as $p_c=0.34457(1)$ \cite{GrassbergerZhang:1996}.
This model belongs to the 2D DP universality class as well.

\begin{figure}
\begin{center}
\includegraphics*[width=0.95\linewidth]{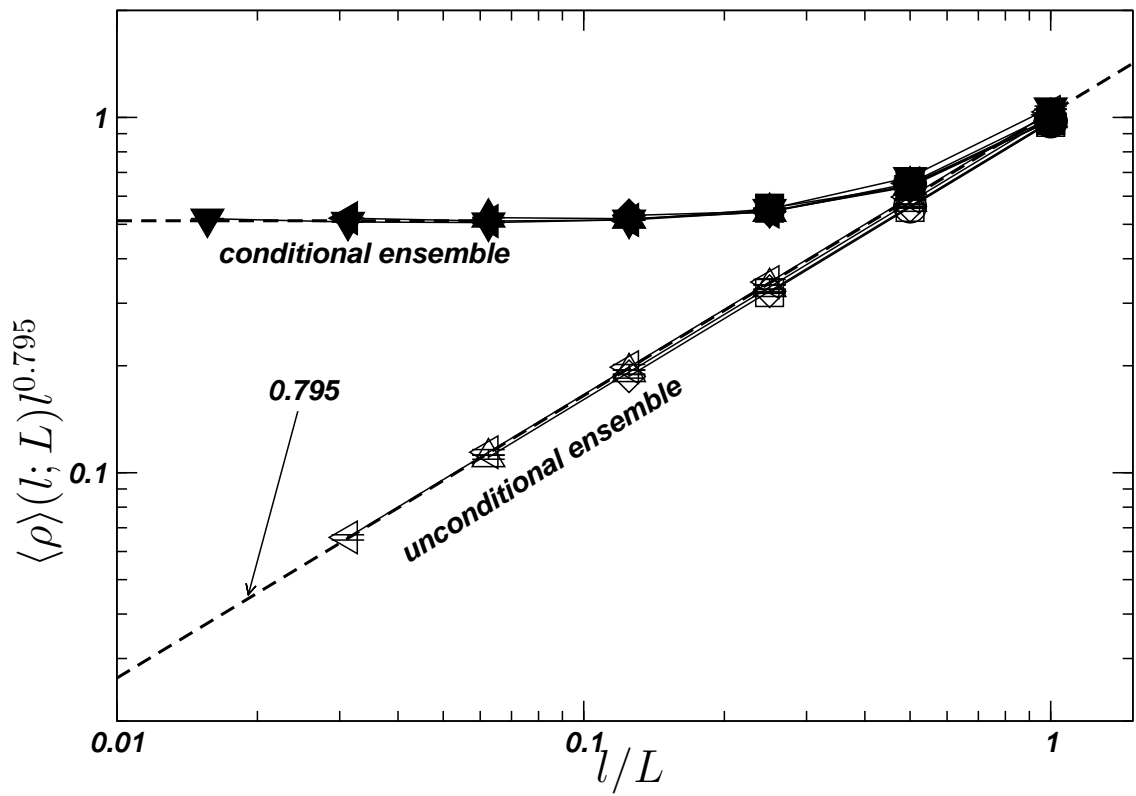}
\end{center}
\caption{\flabel{oparam_scaling}
Data collapse of the order parameter measured in the unconditional (open
symbols) and the conditional ensemble (filled symbols) for the
two-dimensional contact process with system sizes $L=32,64,\ldots,512$
(up to $L=1024$ in the conditional ensemble)
and $l=16,32,\ldots,L$ (data points for the same system size obtained
from the same samples are
connected by lines). The unconditional order parameter collapses
trivially (see text) and is shown for illustration purposes only.
The sloped dashed line has slope $0.795\approx\beta/\nu_\perp$, 
indicating the power-law behaviour of the
scaling function of the unconditional ensemble, $\GC_u$, defined in
\eref{oparam_scaling}.
The conditional
order parameter collapses well under the
scaling ansatz \eref{oparam_scaling}, plotting $\Cave{\activity}(l;L)
l^{\beta/\nu_\perp}$ vs. $l/L$.  The horizontal dashed line is the
likely asymptote of the scaling function $\GC_c$ in the conditional
ensemble. }
\end{figure}

All measurements are taken at the quasi-stationary state: Starting from
random initial configurations with a small but non-vanishing 
activity, the systems evolve according to the rules described above,
until the observables reach a (quasi-)stationary state. For example, in the
two-dimensional CP for $L=256$, the first $10^5$ updates are discarded as
transient. Measurements are taken at constant rate after 
the transient and enter with
the same weight until the system reaches the absorbing state (for
$L=256$ the average lifetime was about $4.8\cdot10^4$) or a maximum time
is reached ($5\cdot 10^5$ for $L=256$). The procedure is repeated until
the statistical error is acceptably small; for example $1.42\cdot10^8$
systems of size $L=256$ were started, of which only about $13\%$ survive
the transient. Statistical errors were estimated by sub-sampling the
ensemble,
which copes even with correlated data. 
In the following, the scaling of the various (block) observables in $l$
and $L$ is analysed.  

\subsection{Order parameter}
The first moment of the activity in the
unconditional ensemble, $\Uave{\activity}(l;L)$, does not vary
in $l$ at all, because of translational invariance: Every (randomly
chosen) site is
equally likely to be active and therefore $\Uave{\activity}(l;L)$ is
constant in $l$.  In order to establish ordinary FSS complying to
\eref{oparam_scaling}, the scaling function $\GC_u(x)$ must necessarily
be a power-law itself, $\GC_u(x) = \GC_u(1) x^{\beta/\nu}$, so that
\begin{equation}
\Uave{\activity}(l;L)=C_u \GC_u(1) L^{-\beta/\nu}
\elabel{Uactivity_scaling}
\end{equation}
otherwise
standard finite size scaling, $\Uave{\activity}(l=L;L)\propto
L^{-\beta/\nu}$, would not be recovered for $l=L$, \ie when a single block
covers the entire system.  Contrary to what is
expected from equilibrium critical phenomena, this scaling function
necessarily vanishes at $0$, \ie $\lim_{x\to0}\GC_u(x)=0$.
\Fref{fig:oparam_scaling} contains a (trivial) data collapse for
$\Uave{\activity}(l;L)$ according to
\eref{oparam_scaling}, namely $\Uave{\activity}(l;L)
l^{\beta/\nu_\perp} \propto (l/L)^{\beta/\nu_\perp}$ 
as a function of $l/L$ for various system sizes
$L$, which illustrates the scaling behaviour.

The conditional order parameter $\Cave{\activity}(l;L)$, on the other
hand, does not suffer from this complication. By construction at least
one site per patch is active,
$\Cave{\activity}(l;L)\ge l^{-d}$, so that
$\lim_{L\to\infty}\Cave{\activity}(l;L)\ge l^{-d}$. The latter limit is
the thermodynamic limit of a density and its existence is the most basic
assumption in statistical mechanics. Provided $l$ is sufficiently large
compared to the fixed lower cutoff $l_0$, so that \eref{oparam_scaling}
applies, the limit implies that $\lim_{x\to0}\GC_c(x)>0$, \ie
$\GC_c(x)$ converges to a non-zero value. This is numerically confirmed
by the data collapse of $\Cave{\activity}(l;L)$ in \Fref{fig:oparam_scaling}.

\subsection{Variance of the order parameter}
\begin{figure}
\vspace*{0.5cm}
\includegraphics*[width=0.95\linewidth]{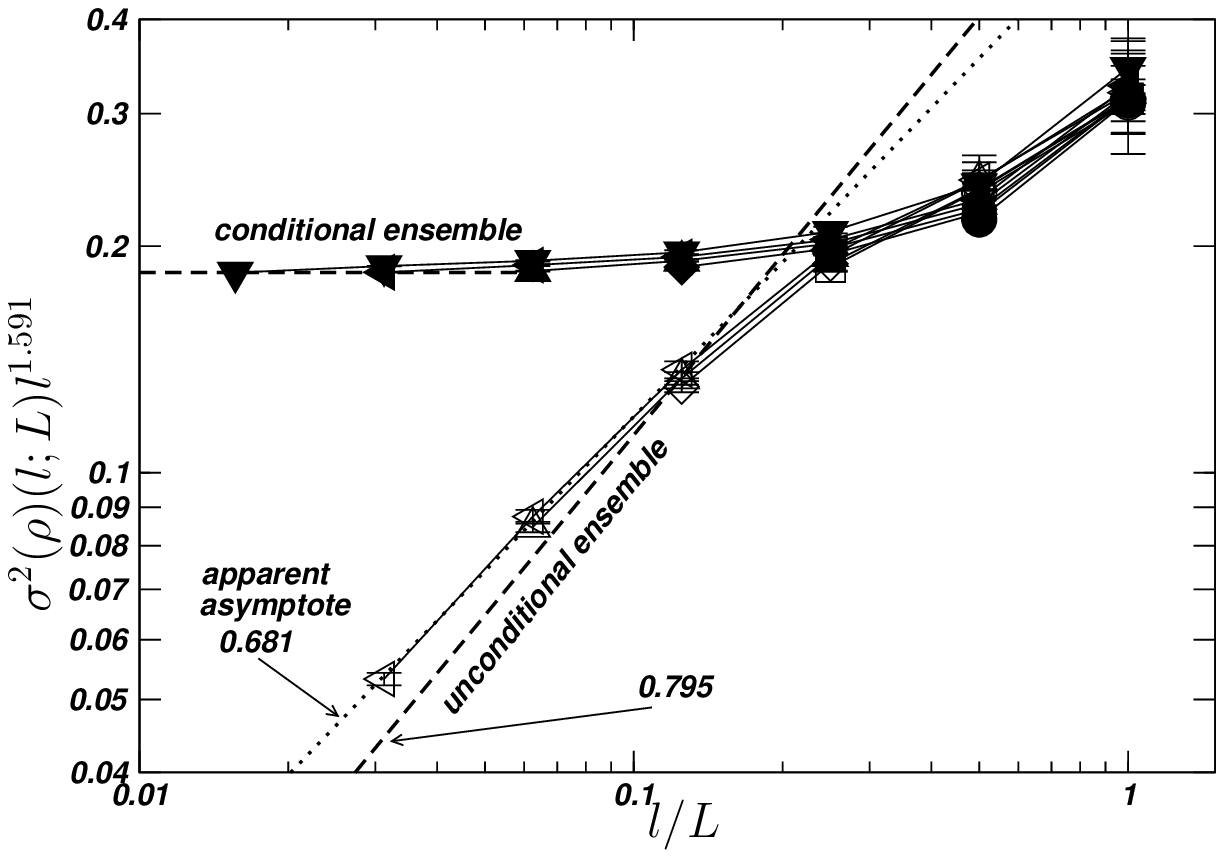}
\vspace*{-0.5cm}
\caption{\flabel{var_box_scaling}
Data collapse of the variance of the order parameter in the
unconditional (open symbols) and the conditional ensemble (filled
symbols), as in \Fref{fig:oparam_scaling}.  The data is shown in rescaled
form, as $\var{\activity}(l;L)l^{2\beta/\nu_\perp}$ vs. $l/L$.  Again, the
horizontal dashed line is the likely asymptote of the scaling function
$\FC_c$ (see \eref{var_box_scaling}) of the conditional ensemble, while
the sloped dashed line has slope $\beta/\nu_\perp\approx0.795$, expected to be
the asymptote of the unconditional ensemble. The dotted line (apparent
asymptote) with slope
$0.681$, however, fits the numerical data very convincingly, indicating
that the asymptotic regime has not yet been reached.  }
\end{figure}

The considerations are similar for the second cumulant,
$\UCvar{\activity}(l;L)$. At $l=1$ the second moment coincides with the
first, so that
$\UCvar{\activity}(l=1;L)=\UCave{\activity}(l=1,L)-\UCave{\activity}^2(l=1,L)$.
In the unconditional ensemble, this quantity scales asymptotically like
$L^{-\beta/\nu_\perp}$ because it is dominated by
$\Uave{\activity}(l=1,L)=\Uave{\activity}(l,L)\propto
L^{-\beta/\nu_\perp}$. On the other hand, for $l=L$ standard finite size
scaling is to be recovered, $\Uvar{\activity}(l=L;L)\propto
L^{-2\beta/\nu_\perp}$, so that the scaling function $\FC_u(x)$ in
\eref{var_box_scaling} must somehow join these two scaling regimes,
which implies $\FC_u(x)=\FC(1) x^{\beta/\nu_\perp}$ for small arguments
$x$.

However, strictly this argument does not apply, because $l=1$ cannot be
expected to be large compared to the lower cutoff $l_0$ (and in fact is
not in the systems studied numerically in this article). On the other
hand, one might argue that
$\Uvar{\activity}(l;L)/\Uvar{\activity}(l=1;L)$ can be expected to
remain finite in the thermodynamic limit. While this is not a necessity,
the alternative would imply a rather exotic behaviour of the variance,
with the dotted line (the ``apparent asymptote'') in
\Fref{fig:var_box_scaling} moving further and further away from
$\Uvar{\activity}(l=1;L)$ (which necessarily scales like
$L^{-\beta/\nu_\perp}$, dashed line in \Fref{fig:var_box_scaling}) with increasing $L$.
If $\Uvar{\activity}(l;L)/\Uvar{\activity}(l=1;L)$ converges and does
not asymptotically vanish in $L$, then $\Uvar{\activity}(l;L)$ inherits the
scaling of $\Uvar{\activity}(l=1;L)$ in $L$ and $\FC_u(x)= \FC_u(1)
x^{\beta/\nu_\perp}$ for small $x$, so that
$\Uvar{\activity}(l;L) = D_u \FC_u(1) l^{-\beta/\nu_\perp}
L^{-\beta/\nu_\perp}$, for small $l/L$. 

Because the scaling function is a power law only
in the asymptote, for intermediate values of $x=l/L$ the apparent
scaling of $\Uvar{\activity}(l;L)$ might produce very different
effective exponents. This can be seen in \Fref{fig:var_box_scaling} where
the slope suggests $\FC_u(x)\propto x^{0.681}$. A direct estimate
of the scaling of $\Uvar{\activity}(l;L)$ in $l$, at a given, fixed $L$
would then suggest $\Uvar{\activity}(l;L)\propto l^{(-2\beta/\nu_\perp+0.681)}$. This is a reminder
that scaling assumptions like \eref{var_box_scaling} can numerically be
verified only by a data collapse. The effective exponent of $0.681$ is
of course not a universal quantity and its deviation from $0.795$ simply indicates that
asymptotia has not been reached.

Not much can be said about the variance in the conditional ensemble.
While the second moment has a lower bound (namely $\Cave{\activity^2}\ge
l^{-d}$), indicating that its scaling function does not vanish in the
limit of small arguments, no lower bound exists for the variance; in
fact $\Cvar{\activity}(l=1,L)=0$ by construction. It is, however,
reasonable to assume that the variance $\Cvar{\activity}(l,L)$ is finite
in the thermodynamic limit for fixed $l$, because by construction every
patch always retains some activity regardless of the system size. In
contrast, in the unconditional ensemble, the moments of the activity
within a finite fraction of patches might vanish for a duration which increases with
increasing system size, because for
samples to continuously contribute to the average only one site
needs to be active
somewhere in the system.
The scaling function $\FC_c(x)$ for the conditional ensemble being
asymptotically finite, 
$\lim_{x\to0}\FC_c(x)>0$,
is in line with the numerical evidence, see
\Fref{fig:var_box_scaling}.

\subsection{Active fraction $\probabAct(l;L)$}
The scaling of the various observables is linked by
\eref{rel_uave_cave}. The fraction of active blocks, $\probabAct(l;L)$,
is given by any ratio of moments taken in the unconditional and the
conditional ensemble, so that based on the first moments, its scaling is
given by
\begin{equation}
\elabel{probabAct_scaling}
\probabAct(l;L)
=\frac{\Uave{\activity^n}(l;L)}{\Cave{\activity^n}(l;L)}
=\left(\frac{L}{l}\right)^{-\beta/\nu} \frac{C_u}{C_c} 
\frac{\GC_u(1)}{\GC_c(l/L)}
\mathrm{\quad for\quad} l>l_0
\end{equation}
see \eref{oparam_scaling} and \eref{Uactivity_scaling}. As can be seen
from \Eref{rel_uave_cave}, if a moment in
the conditional ensemble scales like $l^{-n\beta/\nu}$, in the
unconditional ensemble it will scale
like $l^{-(n-1)\beta/\nu} L^{-\beta/\nu}$. Similarly, if
$\Uvar{\activity} = D_u \FC(1) l^{-\beta/\nu} L^{-\beta/\nu}$ for small
$l/L$, the variance in the conditional ensemble is dominated by a term
proportional to $(l/L)^{-2\beta/\nu}$.

One could consider the fraction of active blocks $\probabAct(l;L)$ as
the coarse-grained order parameter within a real-space renormalisation
group scheme \cite{Huang:1987}, so that $(L/l)^d$ is the number of coarse-grained sites.
The scaling of such an order parameter is proportional to
$(L/l)^{-\beta/\nu}$, consistent with \eref{probabAct_scaling}.

\subsection{Moment ratio $R$}
The numerical results shown in \Fref{fig:oparam_scaling} and
\Fref{fig:var_box_scaling} all are for the two-dimensional contact process.
Based on FSS in equilibrium critical phenomena,
one would normally expect appropriate moment ratios of the (absolute)
order parameter, such as
$\UCave{\activity}^2(l;L)/\UCave{\activity^2}(l;L)$ to be universal
functions of $l/L$ for $l\gg l_0$ and to converge to a finite value for
$l/L\to0$. Using Eqs.~\eref{oparam_scaling} and \eref{var_box_scaling},
the moment ratio 
\begin{equation}
R_\UCsub(l;L) \equiv
\frac{\UCave{\activity}^2(l;L)}{\UCave{\activity^2}(l;L)}
\end{equation}
is universal assuming universality of the scaling functions
and of the amplitude ratios. However, from what has
been said earlier, in the unconditional ensemble, $R_u(l;L)\propto
(l/L)^{\beta/\nu_\perp}$ while in the conditional ensemble $R_c(l;L)$
indeed converges to a
finite value. 

Because $R_c(l;L)$ is universal, different models belonging to the same
universality class, such as the CP and DP, should produce the same
values. That is indeed the case, as shown \Fref{fig:moment_ratio}. For
$l=L$ the moment ratios based on the 
unconditional and conditional ensembles coincide by construction and can be
compared to the value of $0.7543(4)$
\cite{DickmanKamphorstLealDaSilva:1998} (Table VIII, using
$R_c(L;L)=1/(1+K_2/m_1^2)$) found by
Dickman and Kamphorst Leal da Silva (see the dotted line in 
\Fref{fig:moment_ratio}).
The deviation of $R_c(l;L)$ from $R_c(L;L)$ for
decreasing $l/L$ does not mean that the former is not universal, 
just like 
one would generally expect that
the value of the latter depends on various geometrical and topoligical
properties of the lattice, such as its aspect ratio and the type of
boundary condition \cite{PrivmanHohenbergAharony:1991}.

The numerical
situation for $R_u(l;L)$ (not shown) remained somewhat unclear. Only very few
points for $l/L$ close to $1$ seem to overlap within the error. The
system sizes simulated did not allow a firm statement as to whether
$R_u(l;L)$ is actually universal.

\begin{figure}
\begin{center}
\vspace*{0.5cm}
\includegraphics*[width=0.95\linewidth]{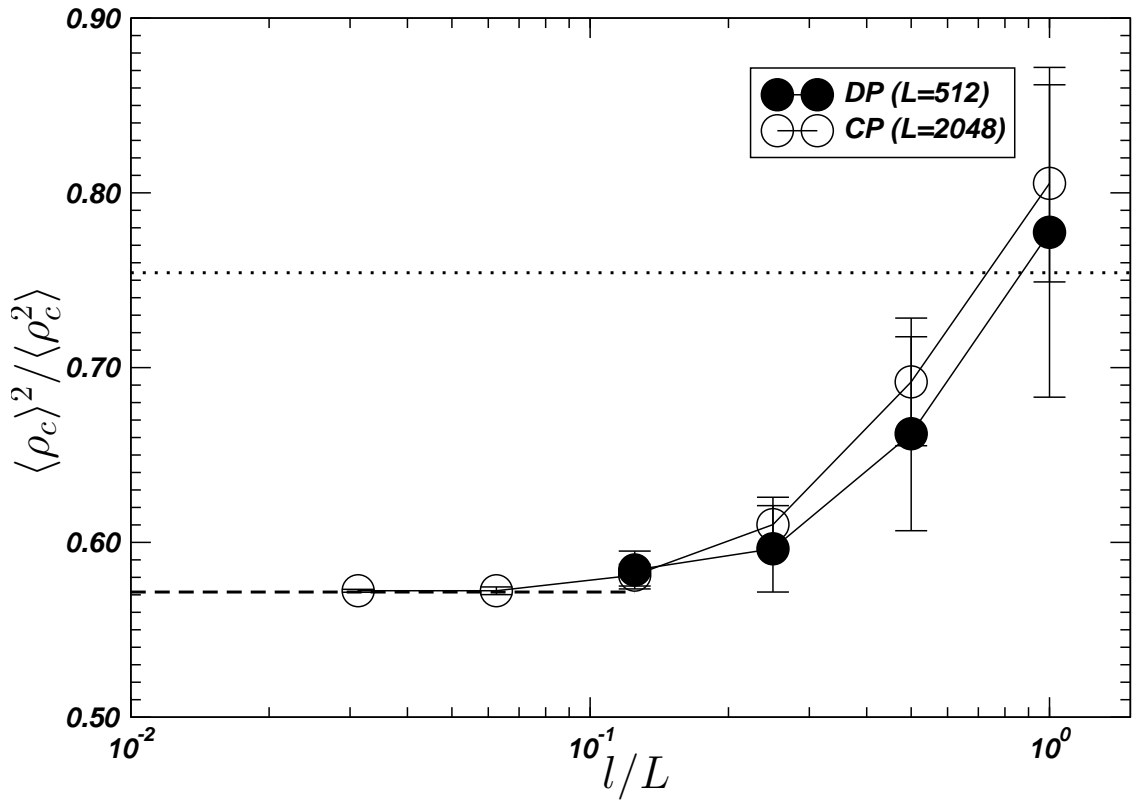}
\vspace*{-0.5cm}
\end{center}
\caption{\flabel{moment_ratio}
The moment ratio $R_c(l;L)=\ave{\activity_c}^2(l;L)/\ave{\activity_c^2}(l;L)$ for the
two-dimensional CP and DP. 
For $l>l_0$ (here $l_0=64$ was needed for satisfactory results) 
the values coincide for both
models, as expected by universality. The dashed line shows the expected
assymptote for small $l/L$, whereas $l/L=1$ is consistent with the
finite size scaling result
$R_c(L;L)=0.7543(4)$ (dotted
line) reported in
\cite{DickmanKamphorstLealDaSilva:1998}.}
\end{figure}

\subsection{Surrogate data}
The proposed method can be put to test and compared to others
using surrogate data, that is
data generated in a computer implementation of, for example, the contact
process, mimicking real-world data as they would be obtained in physical
or biological systems. In contrast to the simulation data used above, such data
consists of a single realisation, as if one was to analyse a satelitte
image or field data.

\begin{figure}
\begin{center}
\includegraphics*[width=0.95\linewidth]{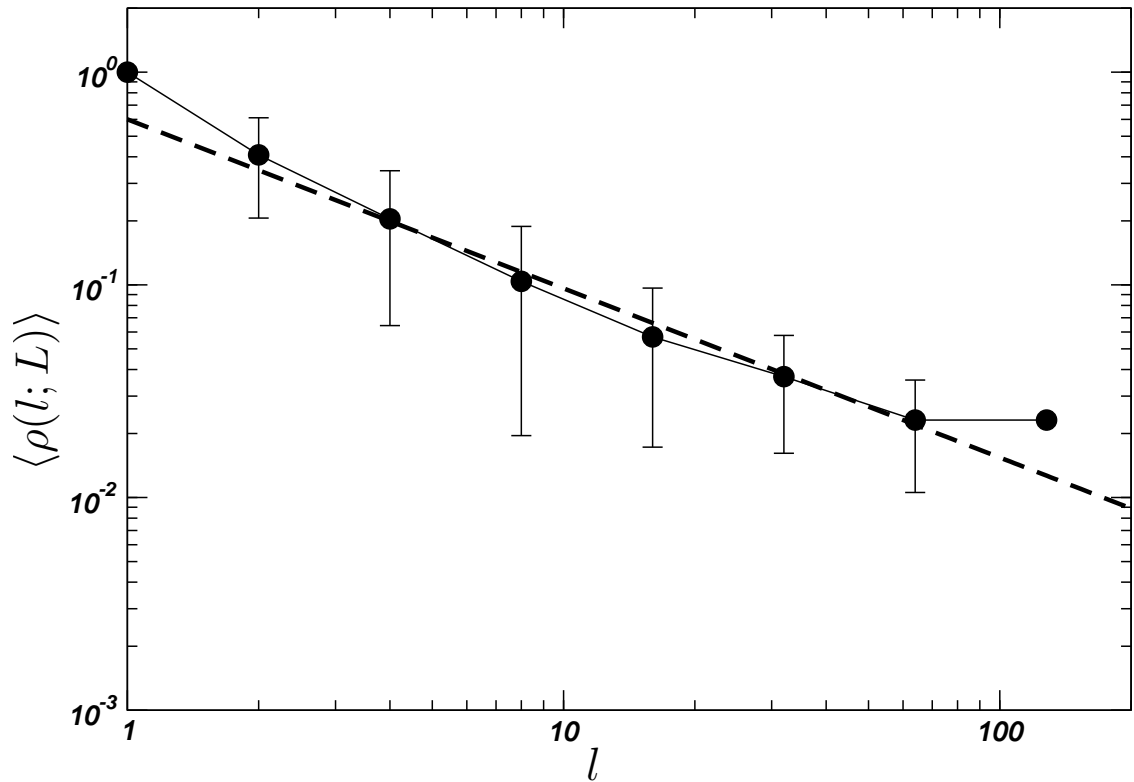}
\end{center}
\caption{\label{pure_cp_boxing_sample_7D.dat_DD.agr}
Conditional box scaling of the activity as measured in a single instance of
the two-dimensional CP as surrogate data. The dashed line indicates the
expected scaling, which fits the data very well.
}
\end{figure}

\Fref{pure_cp_boxing_sample_7D.dat_DD.agr} shows the conditional box
scaling of the activity $\Cave{\activity}(l;L)$ for a single realisation
of a two-dimensional system of linear size $L=128$ as a function of the box size $l$. 
The slope of $\Cave{\activity}(l;L)$ is compared to $l^{-\beta/\nu_\perp}$
(thick dashed line) and
fits very well. No errorbar can be given for $l=L$, as there is only one
such sample, the remaining errorbars are estimated from the variance of
the corresponding sample of size $(L/l)^2$, assuming 
independence.\footnote{Obviously, this amounts to an overestimation of
the number of independent samples and clashes with the assumption of
correlations that give rise the scaling in the first place.}

\begin{figure}
\begin{center}
\includegraphics*[width=0.95\linewidth]{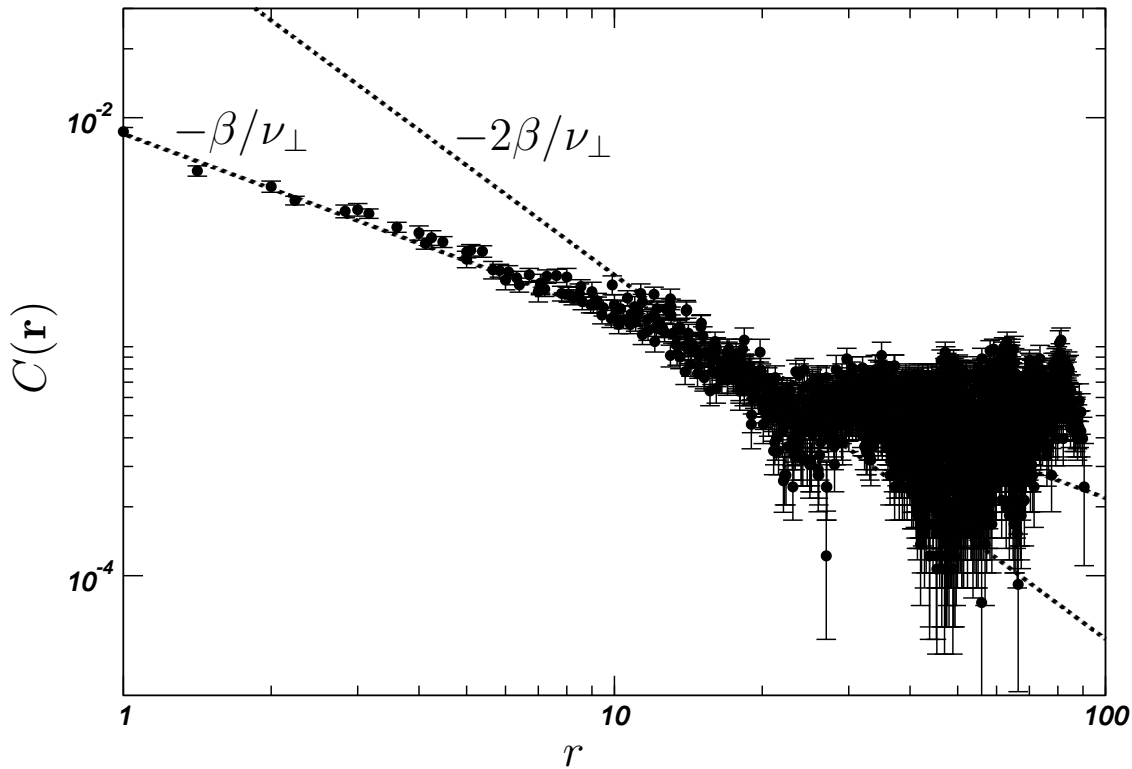}
\end{center}
\caption{\label{pure_cp_boxing_sample_7D.dat_CC.agr}
The two point correlation function (see \Eref{def_corr_fct}) as measured in a single instance of
the two-dimensional CP as surrogate data. The error is underestimated by
assumed independence of the samples.
The dotted lines indicate the
expected scaling. The steeper one corresponds to what is found in
equilibrium critical phenomena, which is superseeded by the shallower
one. This line, with exponent $-\beta/\nu_\perp$ captures the behaviour
for not too large distances quite well.
}
\end{figure}

For comparison, \Fref{pure_cp_boxing_sample_7D.dat_CC.agr} is a double
logarithmic plot of the two point correlation function 
\begin{equation}
C(\rvec) = L^{-2} \sum_{\rvec'} I(\rvec')I(\rvec'+\rvec) -
\left( L^{-2} \sum_{\rvec'} I(\rvec') \right)^2
\elabel{def_corr_fct}
\end{equation}
as a function of the absolute distance $|\rvec|$. Here $I(\rvec)$
indicates occupation of site $\rvec$ and the sums run over all $L^2$
sites $\rvec'$.
It is taken from the
same sample as \Fref{pure_cp_boxing_sample_7D.dat_DD.agr},
using the translational invariance and the eight-fould symmetry of a
square. Nevertheless, the data is
comparatively noisy. 
Na{\"i}ve scaling arguments along the lines of equilibrium phase
transitions \cite{Luebeck:2004} suggest an asymptote
$C(\rvec)\propto |\rvec|^{-d+2-\eta}=|\rvec|^{-2\beta/\nu_\perp}$. 
In \Fref{pure_cp_boxing_sample_7D.dat_CC.agr} it is
shown as a thick dotted line, but this asymptote is compatible
with the data 
only in a narrow, noisy intermediate regime.

The correct scaling behaviour of the correlation function however is
$C(\rvec)\propto |\rvec|^{-\beta/\nu_\perp}$, which is found by
\emph{imposing} that $C(\rvec)/\Uave{\activity}$ does not scale in $L$
\cite{DickmanMartinsdeOliveira:2005} (as it would, for example, in equlibrium
phase transitions, consistent with \cite{GrassbergerdelaTorre:1979}). 
Though still noisy, the data for small $r$ shown in
\Fref{pure_cp_boxing_sample_7D.dat_CC.agr}, indeed is much better compatible
with a scaling exponent of $-\beta/\nu_\perp=-0.795(6)$.

Box scaling of the order parameter is due to the correlations
captured in the two-point correlation function, as can be seen directly
by deriving the variance $\Uvar{\activity}(l;L)$ from $C(\rvec)$, 
\begin{equation}
\Uvar{\activity}(l;L) = l^{-2d} 
\int_{l^d} \ddint{r} \int_{l^d} \ddint{r'} C(\rvec-\rvec') \propto
l^{-\beta/\nu_\perp}
\elabel{var_from_C}
\end{equation}
consistent with 
$\Uvar{\activity}(l;L) \propto
l^{-\beta/\nu_\perp}$ as observed earlier. 
Box scaling therefore can be
regarded as an elegant form of extracting the correlations from the
correlation function. From that point of view, the advantage of box scaling over a
direct investigation of the two-point correlation function is merely down to
its simplicity: Box scaling is well understood theoretically 
and easy to implement in an experiment,
in field work or in a computer simulation.

As a final remark, the quality of the surrogate data shows a strong time
dependence. Starting from a randomly occupied lattice, chosing too short an
equilibration time leads to a lack of correlations, with the system
still being dominated by the independent, random initialisation.
Waiting too
long, on the other hand, means that the system is likely to be low in activity and just
about to die out completely, producing sparse and biased results.

\section{Discussion}
A block scaling analysis could provide a practical method to overcome
the problem of limited availability of data in natural systems and allow
the analysis of a natural system's scaling behaviour without the need of an
entire time series. 
Block scaling effectively is a form of sub-sampling and the analysis
utilises the universal scaling with changing block size, which is
characterised in the present work.
As it turns out, in order to obtain block scaling as
known from equilibrium critical phenomena, one has to use a
conditional ensemble, where moments of the activity in a block enter the
average only conditional to non-vanishing activity.  The situation in
absorbing state phase transitions therefore is very different from what
is expected from equilibrium critical phenomena, where no additional
condition is needed in order to obtain box scaling corresponding to
finite size scaling. 

Box scaling effectively measures the correlations between finite boxes
throughout the system. In (near-)equilibrium systems, \ie systems with a
Hamiltonian, at the critical point, the probability density function of
the order parameter is symmetric around $0$ due to the symmetry of the
Hamiltonian. As contributions with opposite signs cancel, fluctuations
decrease with increasing box size and consequently, the box averaged
(absolute) order parameter and its variance decline.

In absorbing state systems, this mechanism does not exist: The local
order parameter is non-negative and therefore cannot cancel out. In an
unconditional ensemble the order parameter does not change with block
size and fluctuations about the mean are not symmetric. For the order
parameter to vanish, its fluctuations must vanish as well. Nevertheless,
introducing the conditional ensemble restores the behaviour of
(near-)equilibrium systems. 

Within the conditional ensemble,
the scaling functions of the observables considered converge to a
finite value for small arguments, so that the cumulants scale in $l$ with
the expected exponents, $\Cave{\activity}\propto l^{-\beta/\nu_\perp}$ and
$\Cvar{\activity}\propto l^{-2\beta/\nu_\perp}$ for sufficiently small
$l/L$.  Moreover, the moment ratio $\Cave{\activity}^2/\Cave{\activity^2}$
is universal and 
converges to a non-zero, universal value for sufficiently small $l/L$.

This is not the case in the unconditional ensemble: The scaling
functions are power-laws themselves and therefore the moments
display an unconventional scaling. As a consequence, the average activity
is constant in $l$, while its variance scales like $\Uvar{\activity}\propto
l^{-\beta/\nu_\perp} L^{-\beta/\nu_\perp}$ for small $l/L$. In addition,
the moment ratio $\Uave{\activity}^2/\Uave{\activity^2}$ vanishes
asymptotically in small $l/L$.
The unconventional scaling of the unconditional ensemble is due to the
existence of the translational symmetry which is not present in the
conditional ensemble. This symmetry causes the lack of scaling of the
first moment, which is connected to that of all other moments through
\eref{rel_uave_cave}:

The ratio of any unconditional moment and its conditional
counterpart is the fraction of active blocks $\probabAct(l;L)$; this
quantity itself displays universal behaviour in the ratio $l/L$.
Any unconditional moment can be derived from the
corresponding conditional one and vice versa, by multiplying and
dividing by $\probabAct(l;L)$ respectively.

The key advantage of the conditional ensemble is the scaling of
the first moment $\Cave{\activity}$, which is usually the easiest to
determine, carrying the smallest statistical error. In the unconditional
ensemble, the first moment does not scale at all.
Moreover, the moment
ratio $\Cave{\activity}^2/\Cave{\activity^2}$ converges to a finite,
universal value, which again is a quantity with a comparatively small statistical
error. The hope is to use these observables in experimental situations
where the system is guaranteed to be at the critical point.

\ack{
The author thanks an unknown referee for pointing out the asymptotic
scaling of $\Uave{\activity^2}(l=1;L)$, Michael Gastner, Be{\'a}ta Oborny
and Sven L{\"u}beck for interesting discussions and the RCUK for
financial support. 
The author is very grateful to A. Thomas and D. Moore for providing and
supporting the SCAN computing facility.
The present research was carried out in the context
of the ``Borderline Project'' which is supported by the Santa Fe
Institute and the Hungarian National Science Foundation (OTKA K61534).
}

\bibliography{articles,books}

\end{document}